\documentclass [twocolumn,showpacs,nofootinbib,superscriptaddress]{revtex4}
\usepackage{verbatim}
\usepackage{slashed}
\usepackage{epsfig}
\usepackage{feynmf}
\usepackage{graphics}
\usepackage{amsmath}
\usepackage{mathrsfs}
\usepackage{bm}

\begin{document}


\title{Inclusive decays of $\eta_{b}$ into $S$- and $P$-wave charmonium states}


\author{Zhi-Guo He\footnote{Present address: Departament d'Estructura i
Constituents de la Mat\`eria and Institut de Ci\`encies del
Cosmos,Universitat de Barcelona, Diagonal, 647, E-08028 Barcelona,
Catalonia, Spain. }}

\affiliation{Institute of High Energy Physics, Chinese Academy of
Science, P.O. Box 918(4), Beijing, 100049, China.\\ Theoretical
Physics Center for Science Facilities, Beijing, 100049, China.}

\author{Bai-Qing Li}

\affiliation{Department of Physics, Huzhou Teachers College, Huzhou
313000, People's Republic of China}

\date{\today}

\begin{abstract}\vspace{5mm}
Inclusive S- and P-wave charmonium productions in the bottomonium
ground state $\eta_b$ decay are calculated at the leading order in
the strong coupling constant $\alpha_s$ and quarkonium internal
relative velocity $v$ in the framework of the NRQCD factorization
approach. We find the contribution of $\eta_b \rightarrow
\chi_{c_J}+ g\,g$ followed by $\chi_{c_J} \rightarrow J/\psi+
\gamma$ is also very important to the  inclusive $J/\psi$ production
in the $\eta_b$ decays, which maybe helpful to the investigation of
the color-octet mechanism in the inclusive $J/\psi$ production in
the $\eta_b$ decays in the forthcoming  LHCb and SuperB. As a
complementary work, we also study the inclusive production of
$\eta_c$, and $\chi_{cJ}$ in the $\eta_b$ decays, which may help us
understand the X(3940) and X(3872) states.

\end{abstract}



\maketitle

\section{Introduction}

The existence of the spin-singlet state $\eta_{b}$, which is the
ground state of $b\bar{b}$ system, is a solid prediction of the
non-relativistic quark model. Since the discovery of its
spin-triblet partner $\Upsilon$, people have make great efforts to
search for it in various experimental environments, such as in
$e^{+}e^{-}$ collisions at CLEO \cite{Mahmood:2002jd},
in $\gamma\gamma$ collisions at LEP II \cite{Heister:2002if}and in
$p\bar{p}$ collisions at Tevatron\cite{Tseng:2003md}. Unfortunately,
no evident signal was seen in these attempts. Recently, a
significant progress has been achieved by Babar collaboration. After
analysing about $10^{8}$ data, they observed $\eta_{b}$ in the
photon spectrum of $\Upsilon(3S)\to\gamma\eta_{b}$\cite{:2008vj}
with a signal of 10 $\sigma$ significance. They found the hyperfine
$\Upsilon(1S)-\eta_{b}$ mass splitting is
$71.4^{+2.3}_{-3.1}(\text{stat})\pm2.7(\text{syst})$ MeV. Soon
after, it was also seen in
$\Upsilon(2S)\to\gamma\eta_{b}$\cite{:2009pz} by another group in
Babar, and the mass splitting is determined to be
$67.4^{+4.8}_{-4.6}(\text{stat})\pm2.0(\text{syst})$ MeV.

On the theoretical side, considerable works have been done to study
its properties. The mass of $\eta_{b}$ has been predicted by
potential model\cite{Eichten:1994gt}, effective
theory\cite{Brambilla:2001fw} and Lattice QCD \cite{Liao:2001yh}.
And the recent determinations of $\Upsilon(1S)-\eta_{b}$ mass
splitting in the range of $40-60$ MeV
\cite{Ebert:2002pp,Recksiegel:2003fm,Kniehl:2003ap,Gray:2005ur} are
consistent with the Babar's results. Aside from its mass, the
production and decay properties of $\eta_{b}$ have also been
considered. The number of $\eta_{b}$ produced in $e^{+}e^{-}\to
\gamma+\eta_{b}$ at $B$-factories\cite{Chung:2008km} is found to
exceed that produced at LEP II by about an order of magnitude. In
Ref.\cite{Braaten:2000cm}, the authors calculated the production
rates of $\eta_{b}$ at Tevatron Run II and suggested to detect it
through the decay of $\eta_{b}\to J/\psi J/\psi$, while in
Ref.\cite{Maltoni:2004hv}, it was thought that the double $J/\psi$
channel might be overestimated and it was suggested the $\eta_{b}\to
D^{\ast}D^{(\ast)}$ channel to be the most promising channels. An
explicit calculation of $\eta_{b}\to J/\psi J/\psi$ at NLO in
$v^{2}$\cite{Jia:2006rx} and NLO in $\alpha_{s}$\cite{Gong:2008ue}
shown that this branching fraction is of $10^{-8}$ order which is
about four orders of magnitude smaller than that given in
Ref.\cite{Braaten:2000cm}. And the author in
Ref.\cite{Santorelli:2007xg} argued the effect of final state
interactions in $\eta_{b}\to D\bar{D}^{\ast}\to J/\psi J/\psi$ was
also important. Some other exclusive decay modes such as
$\eta_{b}\to \gamma J/\psi$\cite{Hao:2006nf,Gao:2007fv} and
$\eta_{b}$ decays into double charmonia\cite{Braguta:2009df}, and
inclusive decays, e.g. $\eta_{b}\to
c\bar{c}c\bar{c}$\cite{Maltoni:2004hv} and $\eta_{b}\to
J/\psi+X$\cite{Hao:2007rb} have also been taken into account.

However, comparing to the $c\bar{c}$ $^1S_0$ state $\eta_{c}$, our
knowledge about $\eta_{b}$ is quite limited, and doing some further
works is necessary. In this paper, we will systemically study the
inclusive decays of $\eta_{b}$ into $S$- and $P$- wave charmonium
states. The motivations of this work are fourfold. First, in these
processes, the typical energy scale $m_{b}$ in the initial state and
$m_{c}$ in the final state are both much larger than the QCD scale
$\Lambda_{QCD}$, so we can calculate the decay widths perturbatively
and the non-perturbative effect plays a minor role, which will
reduce the theoretical uncertainties. Second, the branching fraction
of the inclusive decay process is much larger than that of the
exclusive process, which makes the test of theoretical prediction
for the inclusive process be more feasible. Third, in
Ref.\cite{Hao:2007rb}, Hao $et$ $al.$ have calculated the branching
ratio of $\eta_{b}\to J/\psi+X$ and found the contribution of the
color octet process $\eta_{b}\to c\bar{c}(^3S_1^{[8]})+g$ is larger
than the one of the color singlet process by about an order. Since
the color-octet process also contributes to P-wave states
$\chi_{cJ}$ production, in which the $\chi_{c1}$ and $\chi_{c2}$ has
about $36\%$ and $20\%$ branching ratio to $J/\psi+\gamma$
respectively, so we expect that the contribution of
$\eta_{b}\to\chi_{cJ}+X$ process followed by $\chi_{cJ}\to
J/\psi+\gamma$ might also be important for inclusive $J/\psi$
production in $\eta_{b}$ decay. Fourth, in recent years, many
charmonium or charmonium-like states have been found at $B$-factory
(see Ref.\cite{Zhu:2007xb,Swanson:2006st,Olsen:2008qw} for a
review). In the further coming LHCb and Super-B, when accumulating
enough data, it might be possible to observe the interesting decays
of $\eta_{b}$ to $X(3940)$ or $X(3872)$ etc..

The $J/\psi$ inclusive production has already been
studied\cite{Hao:2007rb}, and the $J/\psi(\eta_{c},\chi_{cJ})$
production in association with $c\bar{c}$ pair has been discussed in
our previous work\cite{LI}. As important supplements, here we are
going to consider the contribution of  $\eta_{b}\to
\eta_{c}(\chi_{cJ})+gg$ process in the non-relativistic limit at
leading order in $\alpha_{s}$.

\section{NRQCD Factorization Formulism}

Because of the non-relativistic nature of $b\bar{b}$ and $c\bar{c}$
systems, we adopt the non-relativistic QCD (NRQCD) effective
theory\cite{Bodwin:1994jh} to calculate the inclusive decay widths
of $\eta_{b}$ to charmonium states. In NRQCD, the inclusive decay
and production of heavy quarkonium are factorized into the
production of short distance coefficient and the corresponding long
distance distance matrix element. The short distance coefficient can
be calculated perturbatively through the expansion of the QCD
coupling constant $\alpha_{s}$. The non-perturbative matrix element,
which describes the possibility of the $Q\bar{Q}$ pair transforming
into the bound state, is weighted by the relative velocity $v_{Q}$
of the heavy quarks in the heavy meson rest frame.

In the framework of NRQCD, at leading order in $v_{b}$ and $v_c$,
for the $S$-wave heavy quarkonium production and decay, only the the
$Q\bar{Q}$ pair in color-singlet contributes. For $P$-wave
$\chi_{cJ}$ production, the color singlet $P$-wave matrix elements
and color-octet $S$-wave matrix element are both in the same order
of $v_c$. Then the factorization formulas for the processes under
consideration in this work are given by:
 \begin{widetext}
\begin{subequations}
\begin{equation}
\Gamma(\eta_{b}\to
\eta_{c}+gg)=\hat{\Gamma}(b\bar{b}(^1S_0^{[1]})\to
c\bar{c}(^1S_0^{[1]})+X)
\langle\eta_{b}|\mathcal{O}_b(^1S_0^{[1]})|\eta_{b}\rangle\langle\mathcal{O}_c^{\eta_c}(^1S_0^{[1]})\rangle,
\end{equation}
\begin{eqnarray}
\Gamma(\eta_{b}\to
\chi_{cJ}+X)&=&\hat{\Gamma}_1(b\bar{b}(^1S_0^{[1]})\to
c\bar{c}(^3P_J^{[1]})+X)
\langle\eta_{b}|\mathcal{O}_b(^1S_0^{[1]})|\eta_{b}\rangle\langle\mathcal{O}_c^{\chi_{cJ}}(^3P_J^{[1]})\rangle\nonumber\\
&+& \hat{\Gamma}_8(b\bar{b}(^1S_0)\to c\bar{c}(^3S_1^{[8]})+X)
\langle\eta_{b}|\mathcal{O}_b(^1S_0^{[1]})|\eta_{b}\rangle\langle\mathcal{O}_c^{\chi_{cJ}}(^3S_1^{[8]})\rangle,
\end{eqnarray}
\end{subequations}
 \end{widetext}
where the $\hat{\Gamma}$s are the short-distance factors and
$\langle\eta_{b}|\mathcal{O}_b(^1S_0^{[1]})|\eta_{b}\rangle$,
$\langle\mathcal{O}_c^{\eta_c}(^1S_0^{[1]})\rangle$,
$\langle\mathcal{O}_c^{\chi_{cJ}}(^3P_J^{[1]})\rangle$ and
$\langle\mathcal{O}_c^{\chi_{cJ}}(^3S_1^{[8]})\rangle$ are the
long-distance matrix elements. During our calculation of the short
distance coefficients associating with the $P$-wave color-singlet
matrix elements, there will appear infrared divergence. This
divergence will be absorbed into the color octet matrix element
$\langle\mathcal{O}_c^{\chi_{cJ}}(^3S_1^{[8]})\rangle$.

\section{$\eta_{b}\to\eta_{c}+gg$}

We first consider the S-wave $\eta_{c}$ production from $\eta_{b}$
decay. At leading order in $\alpha_{s}$, there are eight Feynman
diagrams for $(b\bar{b}(^1S_0^{[1]})\to c\bar{c}(^1S_0^{[1]})+gg$.
The typical one is shown in Fig.[1a]. The general form of the short
distance coefficient can be expressed as:
\begin{eqnarray}
\hat{\Gamma}((b\bar{b}(^1S_0^{[1]})\to c\bar{c}(^1S_0^{[1]})+gg)=
\frac{\alpha_{s}^{4}}{m_b^{5}}f(r),
\end{eqnarray}
where $r=m_c/m_b$ is a dimensionless parameter. Since there is no
infrared divergence, we calculate $f(r)$ directly using the standard
covariant projection technique\cite{Kuhn:1979bb}. Given
$m_b=4.65$GeV, $m_c=1.5$Gev, we get $f(r)=23.1$. In NRQCD, up to
$v^{4}$ order, the relations between the color singlet matrix
elements and the non-relativistic wave functions are \footnote{For
the color-singlet four-fermion operators, there is a additional
$\frac{1}{2N_c}$ factor compared to those in
Ref.\cite{Bodwin:1994jh} }:
\begin{eqnarray}
\langle\eta_{b}|\mathcal{O}_b(^1S_0^{[1]})|\eta_{b}\rangle=
\frac{1}{4\pi}|R_{1S}^{b}(0)|^2(1+\mathcal{O}(v_b^4)),\nonumber\\
\langle\mathcal{O}_c^{\eta_c}(^1S_0^{[1]})\rangle=
\frac{1}{4\pi}|R_{1S}^{c}(0)|^2(1+\mathcal{O}(v_c^4)).
\end{eqnarray}

\begin{figure}
\begin{center}
\includegraphics[scale=0.6]{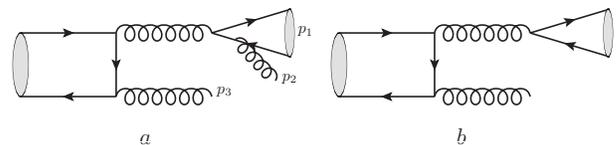}
\caption{Typical Feynman diagrams for the short distance process:
(a) $b\bar{b}[^1S_0,1]\to c\bar{c}[^1S_0^{[1]}(^3P_J^{[1]})]+gg$;
and (b) $b\bar{b}[^1S_0,1]\to c\bar{c}[^3S_1^{[8]}]+g$. }
\end{center}
\end{figure}

In order to compare with our previous work before, we choose the
same numerical values with $m_b=4.65$GeV, $m_c=1.5\mathrm{GeV},
\alpha_{s}=0.22, |R_{1S}^{c}(0)|^2=0.81\mathrm{GeV}^3$, and
$|R_{1S}^{b}(0)|^2=6.477\mathrm{GeV}^3$\cite{Eichten:1995ch}. Then
we get
\begin{equation}
\Gamma(\eta_{b}\to \eta_{c}+gg)=0.83\;\rm{kev}.
\end{equation}

The  total width of $\eta_{b}$ is estimated by using the two gluon
decay, which at leading order in $\alpha_{s}$ and $v_b$ is read to
be:
\begin{equation}
\Gamma_{\rm{Total}}\approx\Gamma(\eta_{b}\to
gg)=\frac{2\alpha_{s}^{2}}{3m_{b}^{2}}|R_{1S}^{b}(0)|^2=9.67\rm{MeV}.
\end{equation}
In our previous work, we got $\Gamma(\eta_{b}\to
\eta_{c}+c\bar{c})\approx0.27\rm{keV}$\cite{LI}. So the branching
ratio of inclusive decay of $\eta_{b}$ into $\eta_{c}$ is
\begin{equation}
\mathrm{Br}(\eta_{b}\to\eta_{c}+X)=1.1\times10^{-4},
\end{equation}
in which the contribution of $gg$ process is about 3 times larger
than that of the $c\bar{c}$ process. The re-scaled energy
distribution curve $d\Gamma/dx_1$ for $\eta_{b}\to\eta_{c}+X$ is
shown in Fig.[2], where $x_1$ is the ratio of $\eta_{c}$ energy
$E_{\eta_{c}}$ to $m_b$.

\begin{figure}
\begin{center}
\includegraphics[scale=0.6]{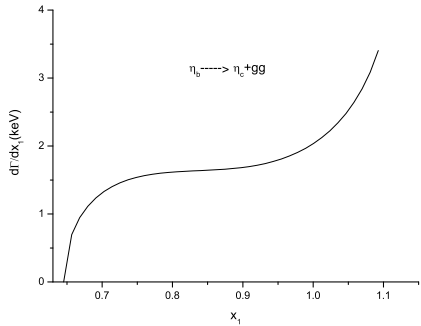}
\caption{The scaled energy distribution of $\eta_c$ for the
$\eta_b\to \eta_c+gg$ process. }
\end{center}
\end{figure}

Recently the $X(3940)$ state was observed by the Belle Collaboration
in the recoiling spectrum of $J/\psi$ in $e^{+}e^{-}$
annihilation\cite{Abe:2007jn}. It is most likely to be a
$\eta_{c}(3S)$ state\cite{Li:2009zu}. In the non-relativistic limit,
the only difference between $\eta_{c}$ and $\eta_{c}(3S)$ is the
value of wave function. If $X(3940)$ is the $\eta_{c}(3S)$ state, we
predict the branching ratio of $X(3940)$ production in $\eta_{b}$
decay to be
\begin{equation}
\mathrm{Br}(\eta_{b}\to X(3940)+X)\simeq0.62\times10^{-4}.
\end{equation}
To obtain the prediction, we have chose
$|R_{3S}^{c}(0)|^2=0.455\mathrm{GeV}^{3}$\cite{Eichten:1995ch} to
take the place of $|R_{1S}^{c}(0)|^2=0.81\mathrm{GeV}^{3}$.

\section{$\eta_{b}\to \chi_{cJ}+gg$}
As mentioned above, the color singlet short distance coefficients
are infrared divergent in full QCD calculation. We will adopt the
dimensional regularization scheme to regularize the divergence. To
absorb the divergence into the color-octet matrix elements
$\langle\mathcal{O}_c^{\chi_{cJ}}(^3S_1^{[8]})\rangle$, we are
necessary to calculate the color-octet short distance coefficient in
$D=4-2\epsilon$ dimensions. The $b\bar{b}(^1S_0^{[1]}\to
c\bar{c}(^3S_1^{[8]})+g$ process includes two Feynman diagram, one
of which is shown in Fig.[1b]. Using the $D$ dimension spin
projector expression\cite{Petrelli:1997ge}, at leading order in
$\alpha_{s}$, the short distance factor is given by
\begin{eqnarray}
\hat{\Gamma}(b\bar{b}(^1S_0)\to
c\bar{c}(^3S_1^{[8]})+g)=\nonumber\\
\frac{(4\pi\alpha_s)^3\mu^{6\epsilon}}{24m_b^5r^3}
\Phi_{2}\frac{(D-2)(D-3)}{(D-1)},
\end{eqnarray}
where $\Phi_{2}=(\frac{\pi}{m_b^2})^{\epsilon}
\frac{\Gamma(1-\epsilon)(1-r^2)}{8\pi\Gamma(2-2\epsilon)}$ is the
2-body phase space in $D$-dimensions.

The calculation of the color-singlet coefficient in full QCD is a
little more complicate. The Feynman diagrams for
$b\bar{b}(^1S_0^{[1]})(P)\to
c\bar{c}(^3P_J^{[1]})(p_1)+g(p_2)g(p_3)$ is the same as those for
$\eta_{c}$ production process. Such $1\to 3$ process can be
described by the following invariants:
\begin{equation}
x_i=\frac{2P\cdot p_i}{M^2},\sum x_{i}=2,
\end{equation}
where $M=2m_b$. In $D=4-2\epsilon$ dimensions, the three-body phase
space is given by
\begin{eqnarray}
&&\mathrm{d}\Phi_{(3)}=KK
((aa+bb-x_2)(x_2+aa-bb))^{-\epsilon}\nonumber\\
&&(1+r^2-x_1)^{-\epsilon}\delta(2-x_1-x_2-x_3)dx_1dx_2dx_3,
\end{eqnarray}
where $r=m_c/m_b$,$aa=\sqrt{x_1^2-4r^2}/2$, $bb=(2-x_1)/2$ and
$KK=\frac{\pi^{2\epsilon}m_b^{2-4\epsilon}}{32\pi^3\Gamma(2-2\epsilon)}$.

In calculating the amplitude, we put the four diagrams with the
gluon carrying a momentum of $p_2$ emitted from the charm quark line
together and label their total amplitude with $M_2$ and  the total
amplitude of the left four diagrams with the gluon carrying a
momentum of $p_3$ emitted from charm quark line is represented by
$M_3$. The total amplitude $M=M_2+M_3$ and
$|M|^2=|M_2|^2+|M_3|^2+2\mathrm{Re}(M_2^{\ast}M_3)$.

As being illustrated in Ref.\cite{Bodwin:1994jh}, for P-wave case
when $p_i(i=2,3)$ goes to zero, there will be singularities in
$M_i$. However, because of the four-momentum conservation, $p_2$ and
$p_3$ can not be soft simultaneously in the phase space. Therefore,
the integration of the interference term
$2\mathrm{Re}(M_2^{\ast}M_3)$ is finite. We could  perform it in
4-dimensions directly. For the symmetry of the two gluons, the
result of phase space integration for $|M_2|^2$ and $|M_3|^2$ are
equal to each other. We only need to calculate one of them. The
total $\hat{\Gamma}_{1}$ then could be written as
\begin{equation}
\hat{\Gamma}_{1}=2\hat{\Gamma}_{M_2}+\hat{\Gamma}_{\mathrm{Int}}.
\end{equation}
where $\Gamma_{M_2}$ and $\Gamma_{\mathrm{Int}}$ are the
contribution related to $|M_2|^2$ and $2\mathrm{Re}(M_2^{\ast}M_3)$
respectively.

We now present how we calculate $\hat{\Gamma}_{M_2}$ in detail. The
denominator of charm-quark propagator in Fig.[1a] is
\begin{equation}
(p_2-p_{\bar{c}})^2-m_c^2=-2p_2\cdot
p_{\bar{c}}\Big{|}_{q_c=0}\propto(1+r^2-x_1-x_2),
\end{equation}
where $p_{\bar{c}}=\frac{p_2}{2}-q_c$ is the momentum of anti-charm
quark and $q_c$ is the relative momentum of $c$ and $\bar{c}$. When
$c\bar{c}$ in $P$-wave configuration, we need to know the first
derivative of the amplitude with respect to $q_c$. Then in the
non-relativistic limit, there will be three kinds of the divergences
in $|M_2|^2$, which are proportional to
$\frac{x_2^2}{(1+r^2-x_1-x_2)^4},\frac{x_2}{(1+r^2-x_1-x_2)^3}$ or
$\frac{1}{(1+r^2-x_1-x_2)}$. These terms will be divergent at point
$(x_1,x_2)=(1+r^2,0)$ which are not easily to be integrated out. We
introduce two new variables $(x_1^{\prime},x_{2}^{\prime})$, defined
by
\begin{equation}
x_1^{\prime}=x_1, x_2^{\prime}=1-\frac{1+r^2-x_1}{x_2}.
\end{equation}
In the variables $x_{1}^{\prime}$ and $x_{2}^{\prime}$, the phase
space is re-expressed as:
\begin{eqnarray}
\mathrm{d}\Phi_{(3)}=\frac{\pi^{2\epsilon}m_b^{2-4\epsilon}}{32\pi^3\Gamma(2-2\epsilon)}
\int^{1+r^2}_{2r} dx_1^{\prime}\int^{1-(bb-aa)}_{1-(bb+aa)}
\frac{dx_2^{\prime}}{(1-x_2^{\prime})^2}\nonumber\\
(1+r^2-x_1)^{1-2\epsilon}((aa+bb-x_2)(\frac{1}{1-x_2^{\prime}}-\frac{1}{bb+aa}))^{-\epsilon},
\nonumber\\
\end{eqnarray}
where $aa=\frac{\sqrt{x_1^{\prime2}-4r^2}}{2}$,
$bb=\frac{(2-x_1^{\prime})}{2}$ and
$x_2=\frac{1+r^2-x_1^{\prime}}{1-x_2^{\prime}}$. And the three
divergence structures are changed to be the form of
$\frac{1}{x_2^{\prime4}}\frac{(1-x_2^{\prime})^2}{(1+r^2-x_1^{\prime})^2}$,
$\frac{1}{x_2^{\prime3}}\frac{(1-x_2^{\prime})^2}{(1+r^2-x_1^{\prime})^2}$
and
$\frac{1}{x_2^{\prime2}}\frac{(1-x_2^{\prime})^2}{(1+r^2-x_1^{\prime})^2}$
respectively, which are all proportional to
$\frac{1}{(1+r^2-x_1^{\prime})^2}$. Then $|M_2|^2$ could be expanded
as
\begin{equation}\label{div}
|M_2|^2=\frac{f_1(1+r^2,x_2^{\prime},\epsilon)}{(1+r^2-x_1)^2}+f_2(x_1^{\prime},x_2^{\prime},\epsilon).
\end{equation}
Accordingly,
\begin{equation}
\hat{\Gamma}_{M_2}=\hat{\Gamma}_{M_2}^{\mathrm{div}}+\hat{\Gamma}_{M_2}^{\mathrm{fin}}.
\end{equation}
Where $\hat{\Gamma}_{M_2}^{\mathrm{fin}}$ is finite and can be
calculate in $D=4$ dimensions. The phase space integration of the
first term in Eq.(\ref{div}) is expressed as
\begin{equation}\label{inte}
\int
\mathrm{d}\Phi_{(3)}\frac{f_1(1+r^2,x_2^{\prime},\epsilon)}{(1+r^2-x_1)^2}=
KK\int^{1+r^2}_{2r}\frac{dx_1^{\prime}g(x_1^{\prime},\epsilon)}{(1+r^2-x_1^{\prime})^{1+2\epsilon}},
\end{equation}
where
\begin{eqnarray}
&&g(x_1^{\prime},\epsilon)=\int^{1-(bb-aa)}_{1-(bb+aa)}
\frac{f_1(1+r^2,x_2^{\prime},\epsilon)}{(1-x_2^{\prime})^2}\nonumber\\
&&((aa+bb-x_2)(\frac{1}{1-x_2^{\prime}}-\frac{1}{bb+aa}))^{-\epsilon}dx_2^{\prime}.
\end{eqnarray}
Furthermore, the integrals in Eq.(\ref{inte}) can be written into
the sum of two terms defined by:
\begin{eqnarray}
&&\int^{1+r^2}_{2r}\frac{dx_1^{\prime}g(x_1^{\prime},\epsilon)}{(1+r^2-x_1^{\prime})^{1+2\epsilon}}\equiv
\int^{1+r^2}_{2r}\frac{dx_1^{\prime}g(1+r^2,\epsilon)}{(1+r^2-x_1^{\prime})^{1+2\epsilon}}+\nonumber\\
&&
\int^{1+r^2}_{2r}\frac{dx_1^{\prime}(g(x_1^{\prime},\epsilon)-g(1+r^2,\epsilon))}{(1+r^2-x_1^{\prime})^{1+2\epsilon}}.
\end{eqnarray}
The first term on the right side includes $\frac{1}{\epsilon}$ pole,
and the second term is finite. Therefore we only need to keep the
$\mathcal{O}(\epsilon)$ contribution when calculating
$g(1+r^2,\epsilon)$ and the second term can evaluated directly by
setting $\epsilon=0$.

Putting Eq.(11) and (16) into together, we get
\begin{equation}
\hat{\Gamma}_{1}=2(\hat{\Gamma}_{M_2}^{\mathrm{div}}+\hat{\Gamma}_{M_2}^{\mathrm{fin}})
+\hat{\Gamma}_{\mathrm{Int}}.
\end{equation}
$\hat{\Gamma}_{M_2}^{\mathrm{div}}$ is calculated analytically, and
$\hat{\Gamma}_{M_2}^{\mathrm{fin}}$ and
$\hat{\Gamma}_{\mathrm{Int}}$ are calculated numerically. For
$J=0,1,2$, the expressions for $\hat{\Gamma}_{M_2}^{\mathrm{div}}$
are
\begin{widetext}
\begin{subequations}
\begin{eqnarray}
&&\hat{\Gamma}_{M_2}^{\mathrm{div}}= \frac{128\left( -1 + r^2
\right)C_AC_F(\alpha_s\pi\mu^{2\epsilon})^{4}KK}{81\,m_b^9\,r^5\,\epsilon}+\nonumber\\&&
  \frac{64C_AC_F\pi^4\alpha_s^{4}KK\left( 4 - 4r^6 +24\left( 1-r^2\left( 3
- 3r^2 + r^4 \right)\right)
     \log (1 - r^2) + 12\left( 1 + r^6 \right) \log (r)\right)}
     {243m_b^9r^5{\left( -1 + r^2 \right) }^2}
 (J=0)
\end{eqnarray}
\begin{eqnarray}
&&\hat{\Gamma}_{M_2}^{\mathrm{div}}= \frac{128\left( -1 + r^2
\right)C_AC_F(\alpha_s\pi\mu^{2\epsilon})^{4}KK}{81\,m_b^9\,r^5\,\epsilon}+
  \frac{128C_AC_F\pi^4\alpha_s^{4}KK}{{243m_b^9r^5{\left( -1 + r^2 \right)
  }^2}}\nonumber\\&&
     \left( 2 - 9r^2 + 9r^4 - 2r^6 + 3\left( 2 - 3r^2 - 3r^4 + 2r^6 \right) \log (r) +
         12\left(1- 3r^2 + 3r^4 - r^6\right) \log (1 - r^2) \right) (J=1)
\end{eqnarray}
\begin{eqnarray}
&&\hat{\Gamma}_{M_2}^{\mathrm{div}}= \frac{128\,\left( -1 + r^2
\right)C_AC_F(\alpha_s\pi\mu^{2\epsilon})^{4}KK}{81m_b^9\,r^5\,\epsilon}
+ \frac{128C_AC_F\pi^4\alpha_s^{4}KK}{1215m_b^9r^5\left( -1 + r^2
\right) ^2}
  \nonumber\\&&\left( 10 - 27r^2
+ 27r^4 - 10r^6 + 3\left( 10 - 9r^2 - 9r^4 + 10r^6 \right) \log (r)
- 60{\left( -1 + r^2 \right) }^3\log (1 - r^2) \right)
 (J=2).
\end{eqnarray}
\end{subequations}
\end{widetext}
The $C_A=3$ and $C_F=4/3$ in above equations are the color factors.
It can be seen that for different $J$ the divergence part of
$\hat{\Gamma}_{M_2}^{\mathrm{div}}$ are the same, which will be
absorbed into the color-octet matrix element. And
$2\hat{\Gamma}_{M_2}^{\mathrm{fin}}+ \hat{\Gamma}_{\mathrm{Int}}$
are
\begin{eqnarray}
2\hat{\Gamma}_{M_2}^{\mathrm{fin}}+
\hat{\Gamma}_{\mathrm{Int}}=\frac{\alpha_s^4}{m_b^7}A_{J}(r) \;
(\mathrm{for}J=0,1,2).
\end{eqnarray}
When $r=1.5/4.65$, we obtain $A_{0}(r)\simeq-9.71\times10^2$,
$A_{1}(r)\simeq-2.66\times10^2$ and $A_{2}(r)\simeq-6.06\times10^2$.

To cancel the infrared divergence of
$\hat{\Gamma}_{M_2}^{\mathrm{div}}$, we also need to take into
account the renormalization of
$\langle\mathcal{O}_c^{\chi_{cJ}}(^3S_1^{[8]})\rangle$. In
$\overline{MS}$ scheme, it is given
by\cite{Bodwin:1994jh,Petrelli:1997ge}
\begin{eqnarray}
\langle\mathcal{O}_c^{\chi_{cJ}}(^3S_1^{[8]})\rangle^{(\Lambda)}=
\langle\mathcal{O}_c^{\chi_{cJ}}(^3S_1^{[8]})\rangle^{\mathrm{(Born)}}-\nonumber\\
\frac{4\alpha_{s}C_F}{3\pi
m_c^2}(\frac{1}{\epsilon}+\log4\pi-\gamma_{E})(\frac{\mu}{\mu_{\Lambda}})^{2\epsilon}
\sum_{J=0}^{2}\langle\mathcal{O}_c^{\chi_{cJ}}(^3P_J^{[1]})\rangle.
\end{eqnarray}

Combining the results in Eq.(8,21,22,23), we finally obtain the
infrared safe expressions for inclusive decay of $\eta_{b}$ into
$\chi_{cJ}(J=0,1,2)$ states
\begin{equation}
\Gamma(\eta_{b}\to \chi_{cJ}+X)=\Gamma_{8}^{J}+\Gamma_{1}^{J},
\end{equation}
where $\Gamma_{8}^{J}$ is
\begin{equation}
\frac{2\pi^2\alpha_s^3(1-r^2)}{9m_b^5r^3}
\langle\eta_{b}|\mathcal{O}_b(^1S_0^{[1]})|\eta_{b}\rangle
\langle\mathcal{O}_c^{\chi_{cJ}}(^3S_1^{[8]})\rangle,
\end{equation}
and $\Gamma_{1}^{J}$ are
\begin{widetext}
\begin{subequations}
\begin{eqnarray}
&&\Gamma_{1}^{0}=\frac{8\pi\alpha_s^4\langle\eta_{b}|\mathcal{O}_b(^1S_0^{[1]})|\eta_{b}\rangle
\langle\mathcal{O}_c^{\chi_{cJ}}(^3P_0^{[1]})\rangle}{243m_b^7r^5(1-r^2)^2}
(12(r^6+1)\log(r)+24(1-3r^2+3r^4-r^6)\log(1-r^2)+\nonumber\\&&2(1-r^2)
((6\log2-5)r^4-4(3\log2-4)r^2+6\log2-5+6(1-r^2)^2\log(\frac{m_b}{\mu_\Lambda}))
+\frac{243r^5(1-r^2)^2A_0(r)}{8\pi}),
\end{eqnarray}
\begin{eqnarray}
&&\Gamma_{1}^{1}=\frac{16\pi\alpha_s^4\langle\eta_{b}|\mathcal{O}_b(^1S_0^{[1]})|\eta_{b}\rangle
\langle\mathcal{O}_c^{\chi_{cJ}}(^3P_0^{[1]})\rangle}{243m_b^7r^5(1-r^2)^2}
(3(2r^6-3r^4-3r^2+2)\log(r)+12(1-r^2)^3\log(1-r^2)+\nonumber\\&&
+(1-r^2)((6\log2-5)r^4+(7-12\log2)r^2+6\log2-5+6(1-r^2)^2\log(\frac{m_b}{\mu_\Lambda}))+
\frac{243r^5(1-r^2)^2A_1(r)}{16\pi}),
\end{eqnarray}
\begin{eqnarray}
&&\Gamma_{1}^{2}=\frac{16\pi\alpha_s^4\langle\eta_{b}|\mathcal{O}_b(^1S_0^{[1]})|\eta_{b}\rangle
\langle\mathcal{O}_c^{\chi_{cJ}}(^3P_0^{[1]})\rangle}{1215m_b^7r^5(1-r^2)^2}
(3(10r^6-9r^4-9r^2+10)\log r+60(1-r^2)^3\log(1-r^2)+\nonumber\\&&
(1-r^2)(5(6\log2-5)r^4+(53-60\log2)r^2+5(6\log2-5)+30(1-r^2)^2\log(\frac{m_b}{\mu_\Lambda}))+
\frac{1215r^5(1-r^2)^2A_1(r)}{16\pi}). \nonumber\\
\end{eqnarray}
\end{subequations}
\end{widetext}
It can be seen that the contribution of $P$-wave color-singlet is
dependent on the factorization scale $\mu_{\Lambda}$. When combining
it with the color-octet $S-$wave contribution, in which the matrix
element also depends on $\mu_{\Lambda}$, the
$\mu_{\Lambda}$-dependence will be canceled.

To give numerical predictions, we also need to know the values of
the long-distance matrix elements. The color octet matrix elements
can be studied in lattice simulations, fitted to experimental data
phenomenologically or determined through some other non-perturbative
ways. Here we determined their numerical values with the help of
operator evolution equations. In the decay process, the solution of
the operator evolution equations are\cite{Bodwin:1994jh}:
\begin{eqnarray}
&&\langle\chi_{cJ}|\mathcal{O}_8(^3S_1;\mu_{\Lambda})|\chi_{cJ}\rangle=
\langle\chi_{cJ}|\mathcal{O}_8(^3S_1;\mu_{\Lambda_0})|\chi_{cJ}\rangle+\nonumber\\
&&\frac{8C_F}{3\beta_0m_c^2}\ln\frac{\alpha_s(\mu_{\Lambda_0})}{\alpha_s(\mu_\Lambda)}
\langle\chi_{cJ}|\mathcal{O}_1(^3P_J)|\chi_{cJ}\rangle,
\end{eqnarray}
where $\beta_0=\frac{11N_c-2N_f}{6}$. We then naively relate the
matrix element of production operator $\mathcal{O}_n^{H}$ and that
of the decay operator $\mathcal{O}_n$ using
\begin{equation}
\langle\mathcal{O}_n^{H}\rangle\approx2J+1\langle
H|\mathcal{O}_n|H\rangle.
\end{equation}
When $\mu_{\Lambda}\gg\mu_{\Lambda_0}$, the evolution term will be
dominated, and the contribution of the initial matrix elements can
be neglected. Since the operator evolution hold only down to scale
$m_c v$, we set the lower bound $\mu_{\Lambda_0}=m_cv$ and choose
$v^2=0.3$. And we set $\mu_{\Lambda}=2m_c$ since the divergence
comes from the soft gluons linked with the $c\bar{c}$ pair. The
$P$-wave color singlet matrix elements can be estimated through
their relations with the first derivative of the non-relativistic
wave function at origin which, in non-relativistic limit, is given
by
\begin{equation}
\langle\mathcal{O}_c^{\chi_{cJ}}(^3P_J^{[1]})\rangle
\approx\frac{3(2J+1)}{4\pi}|R^{\prime}_{c}(0)|^2.
\end{equation}

Setting  $N_f=3$, $\Lambda_{QCD}=390\mathrm{MeV}$ and
$|R^{\prime}_{c}(0)|^2=0.075$ GeV$^5$\cite{Eichten:1995ch}, we
obtain
\begin{equation}
\Gamma(\eta_{b}\to\chi_{cJ}+gg)=(0.17,1.55,1.76)\mathrm{keV}
\;(\mathrm{for}J=0,1,2).
\end{equation}
The $\eta_{b}\to\chi_{cJ}+c\bar{c}$ processes have been considered
in our previous work, in which both the color-singlet and
color-octet contributions have been included but with different
values of the color-octet matrix elements\cite{LI}. If we use the
color-octet matrix elements determined in this work, the results now
become:
\begin{eqnarray}
&&\Gamma(\eta_{b}\to\chi_{cJ}+c\bar{c})=
\nonumber\\&&(4.54,4.21,4.28)\times10^{-2}\mathrm{keV}
\;(\mathrm{for}J=0,1,2),
\end{eqnarray}
which are about an order of magnitude less than the widths of
$\eta_{b}\to \chi_{cJ}+gg$ processes respectively. Including the
contribution of the associate processes, we then predict that the
branching ratios for $\eta_{b}$ inclusive decay into $\chi_{cJ}$ are
\begin{eqnarray}
&&\mathrm{Br}(\eta_{b}\to\chi_{cJ}+X)=\nonumber\\&&
(0.22,1.65,1.87)\times10^{-4} \;(\mathrm{for}J=0,1,2).
\end{eqnarray}

The $X(3872)$ state was discovered in $p\bar{p}$ collisions at
Tevatron\cite{Acosta:2003zx} and $B$ decay at
Belle\cite{Choi:2003ue}. Until now, people have not found an
convincing explanation about it yet. The authors in \cite{Meng:2007cx}
suggest it is a $\chi_{c1}(2P)$ state. If it is a $\chi_{c1}(2P)$
state, we roughly predict
\begin{equation}
\mathrm{Br}(\eta_{b}\to X(3872)+X)=2.25\times10^{-4},
\end{equation}
where we have chose $|R^{\prime}_{c}(0)|^2=0.102$GeV$^5$ and assumed
the ratio between color-singlet and color-octet matrix elements does
not change for $2P$ state.
\begin{equation}
\frac{\langle\mathcal{O}^{\chi_{c1}}_{c}(^3S_1^{[8]})\rangle}{
\langle\mathcal{O}^{\chi_{c1}}_{c}(^3P_1^{[1]})\rangle}
=\frac{\langle\mathcal{O}^{X(3872)}_{c}(^3S_1^{[8]})\rangle}{
\langle\mathcal{O}^{X(3872)}_{c}(^3P_1^{[1]})\rangle}.
\end{equation}

In \cite{Hao:2007rb}, the authors have studied the $\eta_{b}\to
J/\psi+X$ process with $\Gamma(\eta_{b}\to J/\psi+X)=2.29$keV. They
found the contribution of color-octet process $\eta_{b}\to
J/\psi_\mathrm{color-octet}+X$ is more than one order of magnitude
larger than that of the color-singlet contribution. Since
$\chi_{c1}$ and $\chi_{c2}$ could also decay to $J/\psi+\gamma$ with
$\mathrm{Br}(\chi_{c1}\to J/\psi+\gamma)=36\%$ and
$\mathrm{Br}(\chi_{c2}\to J/\psi+\gamma)=20\%$. The branching ratio
of $\chi_{c0}\to J/\psi+\gamma$ is so small that the contribution of
this process can be neglected. Then re-scaling our result by the
values of parameters in Ref\cite{Hao:2007rb}, we find the $\chi_cJ$
feed-down contribution to the decay of $\eta_{b}$ into $J/\psi$ is:
\begin{equation}
\Gamma(\eta_b\to (J/\psi+\gamma)_{\chi_cJ}+X)=0.71\mathrm{keV},
\end{equation}
which is about three times larger than that of color-singlet
process. Therefore in the future experiment, when measuring the
$J/\psi$ production in $\eta_{b}$ decay, the contribution of
$\eta_{b}$ decays into $\chi_{cJ}$ followed by $\chi_{cJ}\to
J/\psi+\gamma$ are also important.

\section{Summary}

In this work, we have studied the inclusive production of charmonium
state $\eta_{c},\chi_{cJ}$ in the decay of ground bottomnium state
$\eta_{b}$ within the framework of NRQCD factorization formula. We
find for the $P$-wave states $\chi_{cJ}$ case, the color-singlet
processes $b\bar{b}(^1S_0^{[1]})\to c\bar{c}(^3P_J^{[1]})+gg$
include infrared divergence. We show that such divergence can be
absorbed into the $S$-wave color-octet matrix element. To give
numerical predictions, we use the potential model results to
determine the color-singlet matrix elements and estimate the
color-singlet matrix elements with the help of operator evolution
equations naively. We find that the branching ratios of $\eta_b$
decays into $\eta_c$ or $\chi_{cJ}$ plus anything are all of
$10^{-4}$ order. Furthermore, we also give the branching ratios of
$\eta_b\to X(3940)+X$ and $\eta_b\to X(3872)+X$, if the $X(3940)$
and $X(3872)$ are the excited $\eta_c(3S)$ and $\chi_{c1}(2P)$
states respectively. In Ref.\cite{Hao:2007rb}, the authors
investigated the color-octet mechanism for $J/\psi$ production in
$\eta_b$ decay, our results show that the $J/\psi$ production from
$\chi_{cJ}$ feed-down is also important, since it is about three
times larger than the direct $J/\psi$ production via color-singlet
channel. These theoretical predictions may not be observed in
experiment for the time being, but will be very helpful to study the
$\eta_b$'s properties in the future experiment such as Super-B.

\section*{\Large{Acknowledgement}}
We would like to thank Yu Jia for helpful discussions. The author
Zhi-Guo He also thanks to the organization of the "Effective Field
Theories in Particle and Nuclear Physics " by KITPC Beijing.





\end{document}